\documentclass[12pt]{iopart}
\usepackage{graphicx}
\usepackage{dcolumn}
\usepackage{bm}
\usepackage{bbm}
\usepackage{multirow}
\usepackage{bbm}
\usepackage{graphicx}
\usepackage{float}
\usepackage{cases}

\begin{document}

\title[J. Phys. A: Math. Theor]{Quantum bit commitment with cheat sensitive binding and approximate sealing}

\author{Yan-Bing Li$^{1,2,3}$, \quad Sheng-Wei Xu$^{2}$, \quad Wei Huang$^{1}$, \quad Zhong-Jie Wan$^{2}$}

\address{$^{1}$State Key Laboratory of Networking and Switching Technology, Beijing University of Posts and Telecommunications, Beijing, 100876, China\\
$^{2}$Beijing Electronic Science and Technology Institute,Beijing,100070,China\\
$^{3}$Department of Electrical Engineering and Computer Science,
Northwestern University, Evanston, Illinois 60208, USA} \ead{liyanbing1981@gmail.com}

\begin{abstract}
This paper proposes a cheat sensitive quantum bit commitment
(CSQBC) scheme based on single photons, in which Alice commits a
bit to Bob. Here, Bob only can cheat the committed bit with
probability close to $0$ with the increasing of used single
photons' amount. And if Alice altered her committed bit after
commitment phase, she will be detected with probability close to
$1$ with the increasing of used single photons' amount. The scheme
is easy to be realized with nowadays technology.
\end{abstract}

\maketitle

\section{ Introduction}

Bit commitment (BC) is a cryptographic task between two
participants, which has a lot of applications to crucial
cryptographic protocols including interactive zero-knowledge
proof~\cite{01,02,03,04}, coin flipping~\cite{05,06,07}, oblivious
transfer~\cite{08,09}, multiparty secure
computation~\cite{10,12,13,14}, and so on.

Generally, BC mainly consists of two phases, $commitment$ phase
and $opening$ phase. In $commitment$ phase, Alice chooses a bit
$b$ ($b =0$ or $1$) which she wants to commit to Bob, and gives
him some encrypted information about the bit, which can not be
decrypted by him before $opening$ phase. Later, in $opening$
phase, Alice announces some information for decrypting $b$ and the
value of $b$. After decryption, Bob obtains an output $b'$. The
commitment would be accepted by Bob if $b' = b$ . Otherwise, the
commitment would be rejected if $b'\neq b$ . Bit commitment must
meet the following needs: \emph{Correctness}. Bob should always
accept with $b' = b$ if both participants are honest.
\emph{Sealing}. Before opening phase, Bob can not know $b$.
\emph{Binding}. Alice can not change $b$'s value after the
commitment phase.

There are several quantum approaches~\cite{05,15} have been
considered to guarantee the unconditional security of quantum BC
(QBC) protocols, such as quantum key distribution (QKD)
protocols~\cite{16,17,18}. Unfortunately, it was concluded that
unconditionally secure QBC can never be achieved in principle,
which was referred to as the Mayers-Lo-Chau (MLC) no-go
theorem~\cite{19,20,21}. Although unconditional secure QBC
protocols are not existent, there are several schemes satisfying
special security models, such as cheat sensitive protocol,
relativistic protocol, have been
proposed~\cite{22,23,24,25,26,27,28,285,29,291,30}. Among them, an
important class is cheat sensitive QBC (CSQBC) which is proposed
by L. Hardy and A. Kent~\cite{22} first. In CSQBC, assuming that
the commitment will eventually be opened, Bob cannot alter the
committed bit after the commitment phase without risking Bob's
detection, and Alice cannot extract information about the
committed bit before the opening phase without risking Bob's
detection as well. In other words, cheat sensitivity means that
all the cheat strategies should be detected with nonzero
probability in the protocol.

In this paper, we propose a variant CSQBC scheme based on single
photons. In the scheme, cheat sensitive is one-way, which is only
available in binding. If Alice alters her committed bit, she will
be detected with probability close to $1$ with the amount's
increasing of used single photons. As for sealing, Bob only can
cheat the committed bit with probability
$\frac{1}{2}+\varepsilon$, where $\varepsilon$ is close to $0$
with the amount's increasing of used single photons. When
$\varepsilon=0$, the one-way CSQBC is more secure than the
two-ways CSQBC as the full sealing is more secure than cheat
sensitive sealing. However, since MLC no-go theorem said
$\varepsilon=0$ is impossible, we only could search for
$\varepsilon\rightarrow0$ in one-way CSQBC.

This paper is organized as follows. Sec. II shows the one-way
CSQBC scheme. In Sec. III, we prove that the scheme is cheat
sensitive in binding and approximate sealing. And the protocol's
practicability is also analyzed. Finally, Sec. IV is a short
conclusion.

\section{\label{sec:level1} The Quantum Bit Commitment Scheme}

In this protocol, Alice will commit a bit $b$ to Bob. Single
photons will be used by them, each of which is prepared as one of
the four states $\{|0\rangle,|1\rangle,|+\rangle,|-\rangle\}$
randomly where $|0 \rangle$ and $|1 \rangle$ are the two
eigenstates of the Pauli operator $\sigma _z$, $|+ \rangle$ and
$|- \rangle$ are the two eigenstates of the Pauli operator $\sigma
_x$. For the cheat sensitive in binding and approximate sealing,
error correcting code (ECC) will be used here. The specific steps
of the protocol are described as follows:

\textbf{[Pre-commitment phase]}

(1) Alice and Bob agree on a ECC $(n, k, d)$-code $C$~\cite{ecc},
which uses $n$ bits codeword to encode $k$ bits word, and the
distance between any two codewords is $d$.

(2) Alice chooses a nonzero random $n$-bit string
$r=(r_{1},r_{2},\cdots,r_{n})$ where $r_{i}\in \{0,1\}$ and
announces it to Bob. Alice uses it to divide all the $n$-bit
codeword $c=(c_{1}c_{2}\cdots c_{n})$ in $C$ into two subsets
$C_{(0)}\equiv \{c\in C|c\odot r=0\}$ and $C_{(1)}\equiv \{c\in
C|c\odot r=1\} $, where $c\odot r\equiv
\bigoplus\limits_{i=1}^{n}c_{i}\wedge r_{i}$.

(3) Bob prepares an ordered $n$ photons sequence
$s=(s_1,s_2,\cdots,s_n )$, in which each $s_i$ is randomly in one
of the four states $(|0\rangle, |1\rangle, |+\rangle, |-\rangle)$.
Then Bob sends the photons sequence $s$ to Alice.

\textbf{[Commitment phase]}

(4) According to the commitment bit $b$, Alice chooses a codeword
$c$ from $C_{(b)}$ randomly.

(5) When $c_{i}=0$, Alice measures the $i$-th photon $s_i$ in
basis $Z$. Else when $c_{i}=1$, Alice measures the $i$-th photon
$s_i$ in basis $X$. Then she obatins the outcomes
$o=(o_1,o_2,\cdots,o_n )$, where $o_i\in
\{|0\rangle,|1\rangle,|+\rangle,|-\rangle\}$.

(6) When $o_i\in \{|0\rangle,|+\rangle\}$, Alice sets $o'_i=0$.
When $o_i\in \{|1\rangle,|-\rangle\}$, Alice sets $o'_i=1$. Then
Alice announces $o'=(o'_1,o'_2,\cdots,o'_t )$ to Bob.

\textbf{[Opening phase]}

(7) Alice announces committed bit $b$, $o$ and $c$ to Bob.

(8) Bob checks whether $o$ is right or not. The rule is that when
$o'_i=0$ (or $1$), it should be $o_i\in \{|0\rangle,|+\rangle\}$
(or $\{|1\rangle,|-\rangle\}$). Then Bob checks whether $c\odot
r=b$ or not. If both of them are right, he accepts the committed
bit. Else, he rejects the committed bit.

\section{\label{sec:level1} Analysis}

In the presented protocol, without considering the noise in the
quantum channels and equipments, Bob will always accept Alice's
committed bit as $c\odot r=b$ when both of them are honest.

However, as a quantum bit commitment protocol, Alice and Bob do
not trust to each other, furthermore, one of them may be dishonest
and perform cheat strategies. So we will analyze the scheme's
security in the following two cases, (1) a dishonest Alice and an
honest Bob, (2) a dishonest Bob and an honest Alice. Generally,
the case that neither Alice nor Bob is honest will not be
considered since it will be a quantum gambling.

And the real-life setting will bring some troubles to the
protocol. In this section, we will analyze the protocol's
practicability following its security analysis.

\subsection{ Cheat sensitive binding}

In the protocol, ECC $(n, k, d)$-code $C$ is used, in which the
distance between any two codewords is $d$. It means that Alice
should change $d$ bits in $c$ if she wants to alter committed bit
$b$ to $b'$, where $b,b'\in \{0,1\}$ and $b\neq b'$. Further,
Alice could use the slyest strategy, in which she first commits a
bit $b''$ other than $0$ or $1$, i.e., she choose a bit string
$c'$ which is contained in neither $C_{(0)}$ nor $C_{(1)}$, and
let the Hamming distance between $c'$ and any one of $C_{(0)}$ and
$C_{(1)}$ be $d/2$. Then she only needs to change $d/2$ bits in
$c'$ to cheat $b=0$ or $b=1$.

When Alice announces $o'$, it means that she had committed
something regardless whether she has measured the photons or not.
In the opening phase, what she should do is to make $o$, and $c$
tally with $o'$ and her wanted $b$. For instance, if she wants to
cheat $b=0$, $c$ should be one in the set $C_{(0)}$. We know that
both of $o_i= |0\rangle$ and $|+\rangle$ ($o_i= |1\rangle$ and
$|-\rangle$) are possible when $o'_i=0$ (or  $o'_i=1$), so $2^n$
different $o$ are legal corresponding to one $o'$. Then the cheat
strategy degenerates to a simpler thing: Bob sends a photon in one
of states $\{|0\rangle,|1\rangle,|+\rangle,|-\rangle\}$ to Alice.
Alice could do anything on it, then she should say whether the
state is in the set $\{|0\rangle,|+\rangle\}$ or
$\{|1\rangle,|-\rangle\}$. If she is right, she could cheat
successfully with probability $1$ as the states in the set are
always legal. But if she is wrong, her cheating will be detected
with probability $1/2$, as she can avoid to be detected when her
announced basis is wrong but be detected with certainty when her
announced basis is right.

Now we analyze how can Alice distinguish the single photon from
the sets $\{|0\rangle,|+\rangle\}$ and $\{|1\rangle,|-\rangle\}$.
Since the photon is always hold in Alice's hand, she would not use
any ancilla states, but measure the photon directly. We suppose
the measurement basis is $\{|r_0\rangle,|r_1\rangle\}$, where
$|r_0\rangle=\cos\theta|0\rangle+\sin\theta|1\rangle$ and
$|r_1\rangle=\sin\theta|0\rangle-\cos\theta|1\rangle$. It should
be that
\begin{subequations}
\label{eq:3}
\begin{eqnarray}
|0\rangle=\cos\theta|r_0\rangle+\sin\theta|r_1\rangle,
\end{eqnarray}
\begin{eqnarray}
|1\rangle=\sin\theta|r_0\rangle-\cos\theta|r_1\rangle,
\end{eqnarray}
\begin{eqnarray}
|+\rangle=\cos(\frac{\pi}{4}-\theta)|r_0\rangle-\sin(\frac{\pi}{4}-\theta)|r_1\rangle
\end{eqnarray}
\begin{eqnarray}
|-\rangle=\sin(\frac{\pi}{4}-\theta)|r_0\rangle+\cos(\frac{\pi}{4}-\theta)|r_1\rangle.
\end{eqnarray}
\end{subequations}

When the photon is $|0\rangle$ or $|1\rangle$, Alice could
distinguish the two sets successfully with probability
$\cos^2\theta$. When the photon is $|+\rangle$ or $|-\rangle$,
Alice would distinguish the two sets successfully with probability
$\cos^2(\frac{\pi}{4}-\theta)$. So the total probability of Alice
distinguishes the two sets successfully is
\begin{equation}
\begin{array}{ll}
P=\frac{\cos^2\theta+\cos^2(\frac{\pi}{4}-\theta)}{2}\\
\hspace{4mm}=\frac{2+\sin2\theta+\cos2\theta}{4}\\
\hspace{4mm}=\frac{2+\sqrt{2}\cos(\frac{\pi}{4}-2\theta)}{4}.
\end{array}
\end{equation}
It should be that $\frac{2-\sqrt{2}}{4}\leq
P\leq\frac{2+\sqrt{2}}{4}$. If Alice distinguishes them
unsuccessfully, Bob will detect the cheating when his basis is
same with what Alice announced. So Alice will be detected with at
least probability $\frac{1-(\frac{2+\sqrt{2}}{4})}{2}$ when she
cheated on one photon. As she must cheat on at least $d/2$
photons, she will be detected with probability
$1-(1-\frac{1-(\frac{2+\sqrt{2}}{4})}{2})^{d/2}=1-(\frac{6+\sqrt{2}}{8})^{d/2}$
for altering the committed bit. With the increasing of $d$, the
probability will be close to $1$. Since $d$ increases with the
increasing of $n$ normally, it means that Alice will be detected
with probability close to $1$ with the amount's increasing of used
single photons if she alters the committed bit.

\subsection{ Approximate sealing}

Before the opening phase, a dishonest Bob might cheat Alice's
committed bit with the states he sent and Alice's announcement.

In fact, without any cheat strategies, a curious Bob could obtain
some information about $o_i$. When the $i$th photon Bob sent is
$|0\rangle$, if Alice said her measurement outcome is in the set
$\{|0\rangle,|+\rangle\}$, he can guess the basis Alice used is
$Z$. Else if Alice said her measurement outcome is in the set
$\{|1\rangle,|-\rangle\}$, he can guess the basis Alice used is
$X$. With this way, he will success with probability $3/4$ to
obtain $o_i$ before the opening phase. However, since the distance
between any two code words in $C_{(0)}$ and $C_{(1)}$ is $d$, Bob
must obtain more than $n-d$ bits to extract valid committed
information. So Bob could cheat successfully with probability
$(\frac{3}{4})^{n-d}$.

Bob has a more sufficient cheat strategy. Instead of sending a
single photon to Alice, Bob could cheat by sending one participle
of an entangle state to Alice. After she measured it, he measures
his participle for analyzing $o_i$. The best thing to him is
obtaining a same state as Alice's, i.e, obtaining a photon in
state $|0\rangle$ (or $|1\rangle$, or $|+\rangle$, or
$|-\rangle$), when $o_i=|0\rangle$ (or $|1\rangle$, or
$|+\rangle$, or $|-\rangle$). Then according to $o'_i$, he
calculates $o_i$ and $c_i$. For these, he should measure the state
which is in one of $\{|0\rangle,|+\rangle\}$ to make sure what
state it is.

The problem of optimal state estimation has been studied in great
detail previously\cite{35}, and in particular the optimal
measurement for discriminating two density operators\cite{36} is
well known. Using the optimal measurement, the maximum probability
that Bob estimates $c_i$ is
\begin{equation}
P^{max}=\frac{1}{2}+\frac{1}{4}Tr|\rho_{|0\rangle}-\rho_{|+\rangle}|=\frac{1}{2}+\frac{\sqrt{2}}{4}
\end{equation},
where
$|\rho_{|0\rangle}-\rho_{|+\rangle}|=\sqrt{(\rho_{|0\rangle}-\rho_{|+\rangle})^\dagger
(\rho_{|0\rangle}-\rho_{|+\rangle})}$ , and
$(\rho_{|0\rangle}-\rho_{|+\rangle})^\dagger$ is Hermitian
conjugate or adjoint of the $(\rho_{|0\rangle}-\rho_{|+\rangle})$
matrix.

So Bob could obtain $c_i$ with success probability
$\frac{1}{2}+\frac{\sqrt{2}}{4}$. Since the distance between any
two code words is $d$, Bob should know more than $n-d$ bits to
obtain valid information. The probability of this case is
$(\frac{1}{2}+\frac{\sqrt{2}}{4})^{n-d}$. Namely, Bob only can
cheat the committed bit with probability
$\frac{1}{2}+\varepsilon$, where $\varepsilon$ is close to $0$
with the increasing of $n-d$. When $\varepsilon\rightarrow 0$,
Bob's cheat strategy almost likes guessing. Since $n-d$ increases
with the increasing of $n$ normally, it means that Bob only can
cheat the committed bit with probability close to $0$ with the
increasing of used single photons' amount.

\subsection{ Practicability}

In the presented protocol, only BB84 states, $X$ and $Y$ bases
measurements are used, all of which can be implemented with
nowadays technology. In QBC, the period between commitment phase
and opening phase may be very long. If quantum states are needed
to be stored during this period, the protocol will be difficult to
realize with nowadays technology. Here, quantum storages are not
needed in the proposed QBC. So compared with some protocols in
which long-time quantum memories are used, our protocol is more
practicable.

Multi-photon is an important problem which has brought some
troubles to practical quantum protocols. Now we analyze its effect
to the presented QBC. We first consider the case happened in $i$th
order. When Bob sends a pulse containing two photons, Alice should
measure one photon in basis $X$, the other in basis $Z$. If the
two outcomes happen to be $\{|0\rangle,|+\rangle\}$ or
$\{|1\rangle,|-\rangle\}$, she can cheat to $c_i=0$ and $c_i=1$
easily by announcing $o'_i=0$ or $o'_i=1$ at step (6) and announce
her wanted $c_i$ at step (7). However, if the two outcomes happen
to be $\{|0\rangle,|-\rangle\}$ or $\{|1\rangle,|+\rangle\}$,
Alice can not perform this cheating. Namely, to one multi-photon,
she could perform the cheating with probability $1/2$. For
cheating successfully, Alice needs to change $d/2$ bits in $c$ at
least. When the multi-photon rate $\eta_m$ is less than
$\frac{d}{2}\times 2\times \frac{1}{n}=\frac{d}{n}$, she could not
cheat successfully. So Bob should set the multi-photon rate of his
source as a small enough value for secure.

The loss and error appearing in quantum channels and devices are
another important problems in practical quantum protocols. Here,
Alice could said some pulse which contains only one photon is
lost. Then she has more chances to cheat. She also could say some
of the attacked bit as error bit. So the loss rate $\eta_l$ and
error rate $\eta_e$ could not be too large. It should be
$\frac{\eta_m}{2}+\eta_l+\eta_e\ll \frac{d}{2n}$.

\section{\label{sec:level1} Conclusion}

To summarize, in this paper, we have dealt with a quantum bit
commitment protocol based on single photons. In our scheme, Alice
commits a value by performing some measurements on the single
photons which are sent from Bob. With the increasing of photons'
amount, Bob only can cheat the committed bit with probability
close to $0$. On the other hand, if Alice alters her committed bit
after commitment phase, she will be detected with probability
close to $1$ with the increasing of photons' amount. It is easy to
be realized with nowadays technology.

\ack This work is supported by NSFC (Grant Nos. 61300181,
61272057, 61202434, 61170270, 61100203, 61121061, 61370188, and
61103210), Beijing Natural Science Foundation (Grant No. 4122054),
Beijing Higher Education Young Elite Teacher Project, China
scholarship council.

\newpage 

\end{document}